# A Novel Chaotic System for Text Encryption Optimized with Genetic Algorithm

Unnikrishnan Menon[1], Atharva Hudlikar[3]

School of Electrical and Electronics Engineering
Vellore Institute of Technology
Vellore, India

Anirudh Rajiv Menon[2]

School of Electronics Engineering
Vellore Institute of Technology
Vellore, India

*Abstract*—**With meteoric developments in communication systems and data storage technologies, the need for secure data transmission is more crucial than ever. The level of security provided by any cryptosystem relies on the sensitivity of the private key, size of the key space as well as the trapdoor function being used. In order to satisfy the aforementioned constraints, there has been a growing interest over the past few years, in studying the behavior of chaotic systems and their applications in various fields such as data encryption due to characteristics like randomness, unpredictability and sensitivity of the generated sequence to the initial value and its parameters. This paper utilizes a novel 2D chaotic function that displays a uniform bifurcation over a large range of parameters and exhibits high levels of chaotic behavior to generate a random sequence that is used to encrypt the input data. The proposed method uses a genetic algorithm to optimize the parameters of the map to enhance security for any given textual data. Various analyses demonstrate an adequately large key space and the existence of multiple global optima indicating the necessity of the proposed system and the security provided by it.**

## I. INTRODUCTION

The Internet is utilized primarily for transmission of data. However, while we take advantage of the Internet's capabilities, unauthorized individuals have the chance to intercept our information and then visit, copy, and destroy it. Therefore, the security and protection of data and information becomes a hot problem studied by experts and enthusiasts alike.

Many encryption systems have gained popularity and momentum over time that have proven to be effective for secure transmission of data. Around 1974, an IBM team developed the Data Encryption Standard (DES) and it was adopted as a national standard in 1977 [1]. Since that time, many cryptanalysts have attempted to find shortcuts for breaking the system. [2]. In 2001, as the outcome of a public competition, Rijndael was announced as the Advanced Encryption Standard (AES) by the US National Institute of Standards and Technology (NIST). Today, the AES is one of the most widely used encryption primitives [3]. Many such algorithms have been used in the past few decades. However, when these algorithms were utilized years ago, the digital technologies were quite different from now and the magnitude of the challenges was lower [4]. For instance, AES suffer from some drawbacks such as, long encryption and decryption time, and patterns appearance in the ciphered image [5].

The theory of dynamical systems and chaos have shown significant scope for research and applications in the field of cryptography [6]. The property of high sensitivity exhibited by a chaotic system with respect to its initial conditions and parameters implies strong cryptographic qualities, and its random-like behavior and unstable orbits with long periods, are quite beneficial to cryptosystems making them resilient to brute force attacks.

Chaos-based ciphers have shown exceptional properties in aspects of security, complexity, speed, and computing power. This is because chaotic maps have many attributes that translate to an efficient cryptographic property. They display an ergodic nature, which means that these systems cannot be broken down to a simplified expression. This adds to the confusion factor of the encryption system. They are also sensitive to initial control parameters. This ensures that the plaintext and/or secret key cannot be obtained easily. A small deviation in the input can cause a large change in the output. Chaotic maps exhibit structure complexity and deterministic dynamics, which allows the cryptographic process to be simple, yet allow for a highly complex encryption and a pseudo-random behavior, respectively [7].

This paper utilizes and proposes a 2D hybrid chaotic map with desirable properties, such as those mentioned above, to generate pseudo-random sequences for text encryption purposes. The proposed map is extremely sensitive to the values of the parameters (taken as the key for encryption) and will therefore return vastly different sequences for minutely dissimilar keys. Hence there arises a need to find optimal key values for a given plaintext.

This problem can be tackled using Genetic Algorithms. As compared to the traditional optimization methods, Genetic Algorithms are robust, global and can be applied generally without making any domain-specific changes. It can be used not only for general, but also for indifferent and unconventional optimization problems [8, 9].

Many genetic algorithm models have been introduced by researchers largely working from an experimental perspective. Most of these studies are application oriented and are typically interested in using them as optimization tools. Researchers have also been improving systems by hybridizing them with genetic algorithms. For instance, a heuristic modified method based on the genetic algorithm for solving constrained optimization problems was introduced in 2009 [10].

In its most generic usage, Genetic Algorithms work by creating a population of agents at every iteration with randomly assigned parameters. These agents are then evaluated based on their performance in the given environment and the best performing agents are carried forward to the next generation of agents. These algorithms encode a potential solution to a specific problem on a simple chromosome-like data structure and apply recombination operators to these structures to preserve critical information. Genetic algorithms are often viewed as function optimizers, although the range of problems to which they have been applied is quite broad [11,12].

This paper explores how genetic algorithms can optimize the key that is used by the proposed chaotic map to generate a pseudo random sequence, favorable for the encryption of a given plaintext.

## II. OPTIMIZATION USING GENETIC ALGORITHM

Genetic algorithms are random heuristic search operations that are developed to imitate the mechanics of natural selection and genetics.

These are population-based search algorithms in which the individuals in the population represent samples from the set of all possibilities.

Genetic algorithms operate on string structures, analogous to biological systems, which are evolving in time according to the rule of survival of the fittest by using a randomized yet structured information exchange. The success of the winning individuals is normally dependent on their genes and is calculated using a fitness function that determines how well an individual performed from the moment it spawned, till termination. A percentage of the best individuals are then chosen which show the most promising fitness values. These individuals are then made to reproduce with each other to spread their genes in the subsequent generations of offspring. The individuals evolve over time to form even better gene variants by sharing and mixing their information about the domain of operation. The genetic algorithm simulates this process and calculates the optimum of objective functions [13].

This paper presents a new functional genetic algorithm optimization methodology that is applied to find out the best set of parameters for the proposed $2D$ chaotic map during the encryption process. This optimization technique ensures that, given the plaintext, the secret key generated will be the one that always ensures maximum confusion and diffusion attributes in the ciphertext thereby enhancing the security of confidential data. In the proposed chaotic map, the coefficients '*a*' and '*b*' need to be fine-tuned dynamically depending on the plaintext. Thus, these 2 coefficients of the chaotic map serve as chromosomes in the genetic algorithm.

The genetic algorithm implemented mainly involves the following 4 major steps:

### A. Evaluation of Fitness

This is done using an objective function that summarizes, as a single figure of merit, how close a given design solution is to achieving the set target. This figure directly indicates how well an individual has performed in the current generation with respect to other individuals belonging to the same generation. The individuals are evolved till a point where certain desirable threshold conditions pertaining to fitness are met.

### B. Selection

Fitness Proportionate Selection is applied to the parent generation. Every individual can become a parent with a probability which is proportional to its fitness. This implies that individuals with larger fitness values have a higher chance of mating and propagating their superior features to the next generation. This strategy applies a selection pressure to the more fit individuals in the population, evolving better off spring over time. The proposed algorithm considers only the top 20% of the individuals in a generation and rejects the rest.

### C. Crossover

This step is analogous to biological reproduction. Parts of the selected parents' chromosomes are copied and pasted to generate one or more offspring. A random crossover point is selected and sections of the two parents from left and right of the crossover point are swapped to get new off-springs.

### D. Mutation

The mutation operation in a genetic algorithm is mainly responsible for exploration of search space. Certain chromosomes in the offspring's DNA are chosen and randomly replaced with variations. This ensures that the new offspring produced are not simply a mere reflection of their previous generation but rather have their own unique features that could potentially lead to better fitness values. In this case, mutation operation is carried out by tweaking the values of the coefficients '*a*' and '*b*' using a relatively small step value which may be either positive or negative.

The evolutionary process terminates when a generation of individuals carries majority of genomes satisfying the desired threshold.

## III. ASSOCIATED CHAOTIC FUNCTIONS

The following maps have played a key role in the conceptualization of the proposed $2D$ chaotic system:

### A. Circle Map

This is a one-dimensional chaotic map that, for certain values of $\Omega$ and $K$, behaves in a chaotic manner [14]. This can mathematically be expressed as:

$$\theta_{n+1} = \theta_n + \Omega - \frac{K}{2\pi}\sin\big(2\pi(\theta_n)\big) \ mod \ 1 \qquad (1)$$

The circle map is an integral part of the proposed chaotic system whose initial conditions and parameter ranges have been defined in Section $IV(c)$.

## B. Hénon Map

In 1976, Hénon published literature detailing a $2D$ chaotic map which could be used as a reduced approach to study the dynamics of the Lorenz system [15, 16]. The Hénon map is mathematically defined by equations (2) and (3):

$$x_{n+1} = 1 - ax_n + y_n \quad (2)$$

$$y_{n+1} = bx_n \quad (3)$$

## IV. Proposed Chaotic Map

The proposed hybrid chaotic map is a $2D$ map which combines 2 different chaotic maps non-linearly. The maps used are Circle Map and Hénon Map. Equations (4) and (5) define this system as given below:

$$x(i+1) = C(a, b, y(i)) \bmod 1 \quad (4)$$

$$y(i+1) = H(a, x(i)) \quad (5)$$

where $C(a, b, y(i))$ is the Circle Map as a function of $a$ and $b$ (the two parameters of the map), $H(a, x(i))$ is the chaotic dimension of the Hénon Map as a function of $a$. The system equation hence becomes:

$$x(i+1) = x(i) + d + (a \sin(2\pi y(i))) \bmod 1 \quad (6)$$

$$y(i+1) = 1 - ax(i)^2 + y(i) \quad (7)$$

### A. Bifurcation Diagram

A bifurcation diagram depicts the values approached asymptotically by a system with respect to its parameters [17]. The map shows the distribution of values taken by the system over a range of the parameters across numerous iterations.

The bifurcation diagram for the proposed system between dimension $x$ and parameter $a$ is shown:

Additionally, a bifurcation diagram was generated between dimension $x$ and the second parameter $b$.

Fig. 1 and 2 show that for all depicted points of $a$ and $b$, the map approaches almost all values in the normalized range, resulting in densely populated bifurcation diagrams. Thus, for $a$ and $b$ values taken over a large range, a uniform chaos is exhibited by the map. From the bifurcation diagrams, the ranges of $a$ and $b$ useful for the purpose of this paper are $(1,4)$ and $(0,4)$ respectively.

### B. Lyapunov Exponent

Lyapunov exponents are used to check whether a system exhibits chaotic behavior. Equation (8) defines a mathematical expression for Lyapunov Exponent and is used to visualize the rate of divergence of a pair of orbits that were initially infinitesimally close [18]. It also infers a sensitivity to a variation in the initial conditions.

$$L_y(f(x)) = \lim_{n \to \infty} \left(\frac{1}{n}\right) \sum_{i=0}^{n-1} \ln |f_i'(x)| \quad (8)$$

where $f_i'(x)$ is the derivative of the $i^{th}$ iterate $f_i(x)$.

A constantly positive value of Lyapunov exponent implies an unstable or chaotic orbit. For the proposed $2D$ chaotic system, $L_x$ and $L_y$ are the Lyapunov exponents for the $x$ and $y$ dimensions, respectively. A system is chaotic if at least one of the Lyapunov exponents remains positive.

For the proposed map, both dimensions exhibit positive Lyapunov exponents across 2000 iterations with values around 4 for both dimensions implying that both dimensions of the map are chaotic in nature as shown in Fig. 3. The first 500 iterations have been ignored to avoid the influence of initial state.

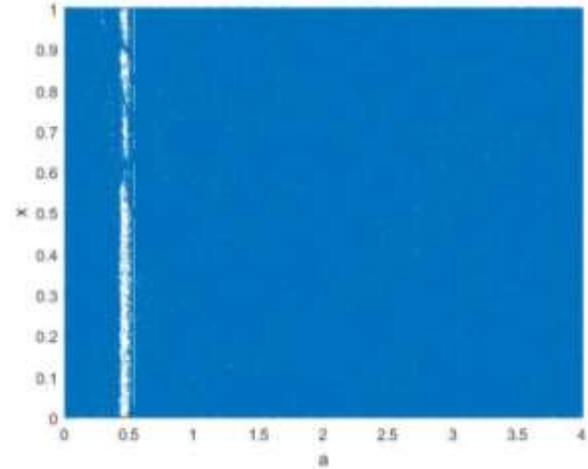

Fig. 1. Bifurcation Diagram of $x$ Dimension vs. Coefficient $a$.

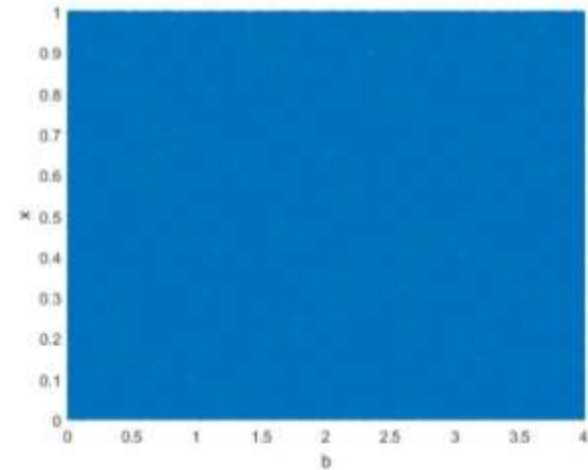

Fig. 2. Bifurcation Diagram of $x$ Dimension vs. Coefficient $b$.

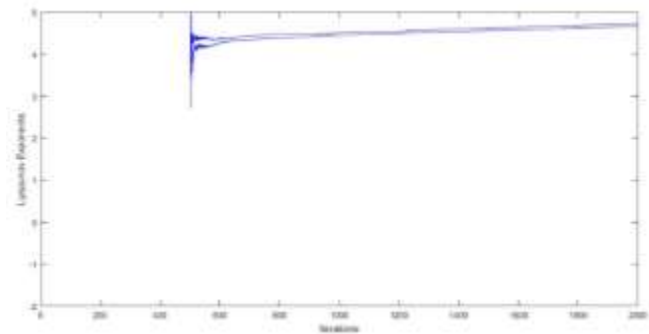

Fig. 3. Lyapunov Exponents for $x$ and $y$ Dimensions.

*C. Selecting Initial Value of Chaotic Map*

The initial $x$ and $y$ dimension values $x_0$ and $y_0$ respectively, of the chaotic map have been set as follows:

$$x_0 = \frac{P_{avg}}{P_{max}} \qquad (9)$$

$$y_0 = 1 - x_0 \qquad (10)$$

where the average and sum of ASCII values of plaintext are represented by $P_{avg}$ and $P_{max}$.

This allows for a different initial value for the $x$ and $y$ dimensions and hence a completely different pseudo random sequence for every plaintext.

V. PROPOSED ALGORITHM

*A. Key Generation*

The key to be generated is the set of parameters (a and b) required by the chaotic map. The key is optimized for the given confidential data using a genetic algorithm Fig. 4.

*1)* An initial population of $(a, b)$ pairs is generated with each pair having random a and b values selected from the ranges defined in Section $IV$.

*2)* Encryption of plaintext is done with each of these pairs (encryption procedure explained in section $V(b)$).

*3)* The fitness function for each (a,b) pair is calculated using the Jaccard index of similarity [19]. It is a statistic used for gauging the similarity between 2 sets and is defined as the size of the intersection divided by the size of the union of the sample sets as shown in equation (11):

$$J(A, B) = \frac{A \cap B}{A \cup B} \qquad (11)$$

The above equation returns a score from $0 - 100$. Where 0 means no similarity at all and 100 implies that both sets are the same. The fitness function taken for the proposed algorithm is:

$$F(genome) = 100 - J(P, E) \qquad (12)$$

where F is the fitness function of the particular genome, P is the set of ascii values of the plain text and E is the ascii set of the encrypted cipher text obtained on encrypting the original text with the a and b values of that genome.

*4)* Once the fitness of each key pair in the population is calculated, the top 20% of the population are promoted (selection).

*5)* A pair of offsprings are generated for every set of parents randomly selected from the promoted/selected set,

where the a and b values of both parents are used to form different combinations of possible key values resulting in the formation of offsprings.

*6)* The selected parents and offspring are then made to undergo mutation where their $(a, b)$ are modified in small random steps with a probability of 0.1

*7)* The mutated set now becomes the next generation.

*8)* The process is repeated for every subsequent generation till a desired fitness value of at least 95% is observed for more than 50% of the individuals belonging to that generation.

*9)* The optimal key is the genome with the best fitness function once the threshold conditions have been met.

*10)* The final encryption is done with the optimised $(a, b)$ values.

*B. Encryption*

*1)* The initial values of the $x$ and $y$ dimensions are taken as shown in section $IV(c)$.

*2)* For a given $(a, b)$ pair, pseudorandom sequences of floating point values, for each dimension of the map i.e. $x$ and $y$, of length equal to size of plaintext are obtained by iterating the chaotic map.

*3)* The arrays of floating point values are each sorted in decreasing order.

*4)* Each element in the original arrays is replaced with the index at which that floating point element appears in the corresponding sorted arrays. This gives rise to two pseudo random arrays of integer values, $S_x$ and $S_y$.

*5)* The final key array will contain values picked from $S_y$ occurring at the indices represented by the values of $S_x$.

*6)* The ciphertext is obtained by performing bitwise-XOR operation between the pseudorandom sequence and the plaintext.

*C. Decryption*

*1)* The key required for decryption consists of the optimized $(a, b)$ along with the initial $x$ and $y$ values used in the encryption phase.

*2)* The final pseudorandom array is generated using the same procedure followed in the encryption step using the aforementioned key.

*3)* The plaintext is retrieved by performing bitwise-XOR operation between the pseudorandom sequence obtained in the previous step and the ciphertext.

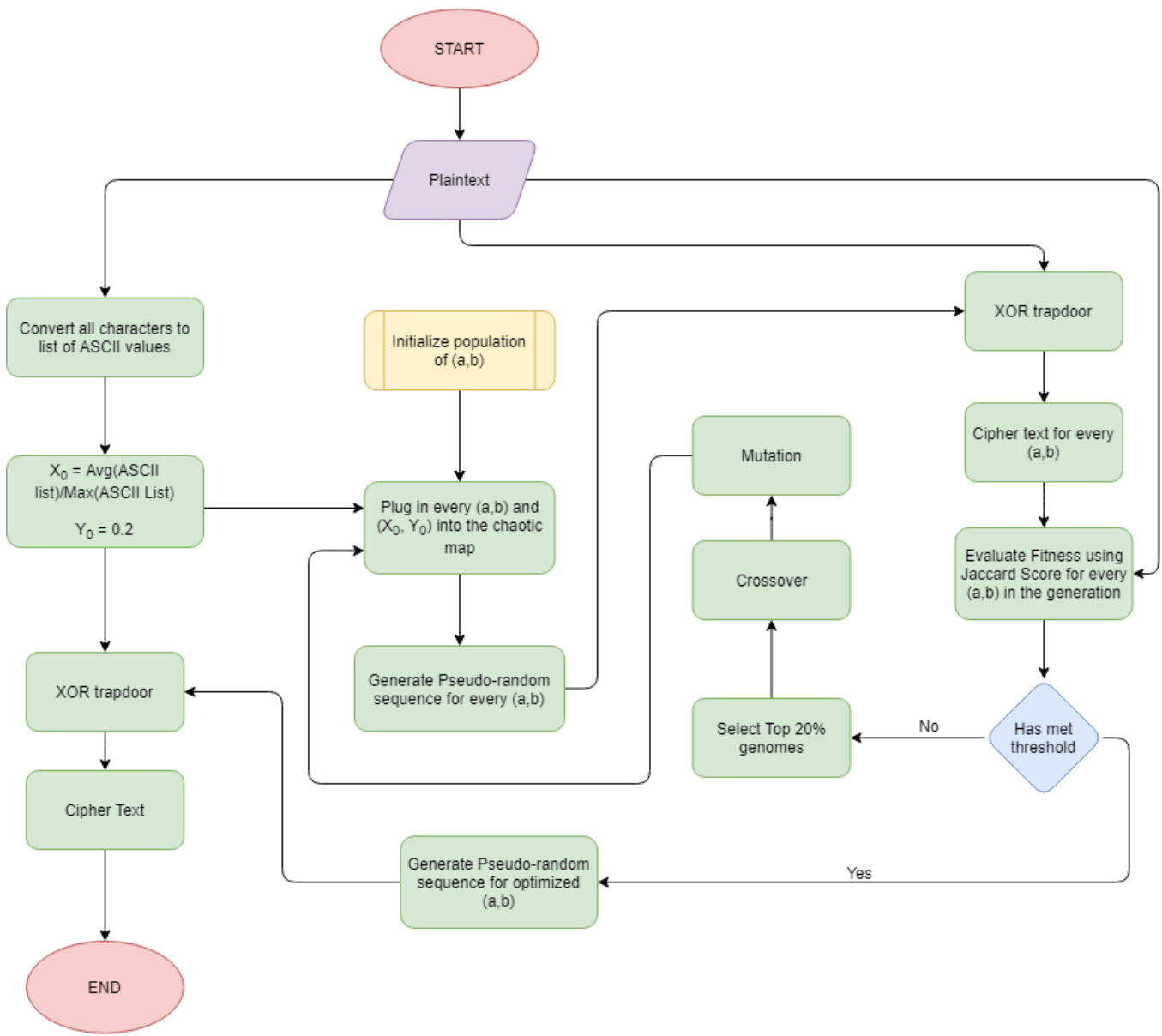

Fig. 4. Flow Diagram of Proposed Algorithm.

## VI. EXPERIMENTAL ANALYSIS

The proposed algorithm was evolved and tested upon multiple plaintext data samples of varying lengths.

The genetic algorithm was deployed for each sample plaintext. The different possibilities of the coefficients $(a, b)$ along with their corresponding fitness values for every genome spawning across all generations were recorded.

For visualization, $2D$ plots of the data were prepared, where the coefficients $a$ and $b$ were plotted on the $x$ and $y$ axes, respectively. The fitness scores of every agent have been graphed as a color scale plot.

Fig. 5 shows that there exist a variety of $(a, b)$ pairs with differing fitness values. While the average fitness demonstrated by an $(a, b)$ pair increases with the length of the plaintext, there clearly exists a set of $(a, b)$ pairs which show maximum fitness scores. This implies that for smaller plaintexts, an extremely minute alteration from the optimum in the values of $(a, b)$ can lead to subpar encryption, thus reinforcing the proposed approach.

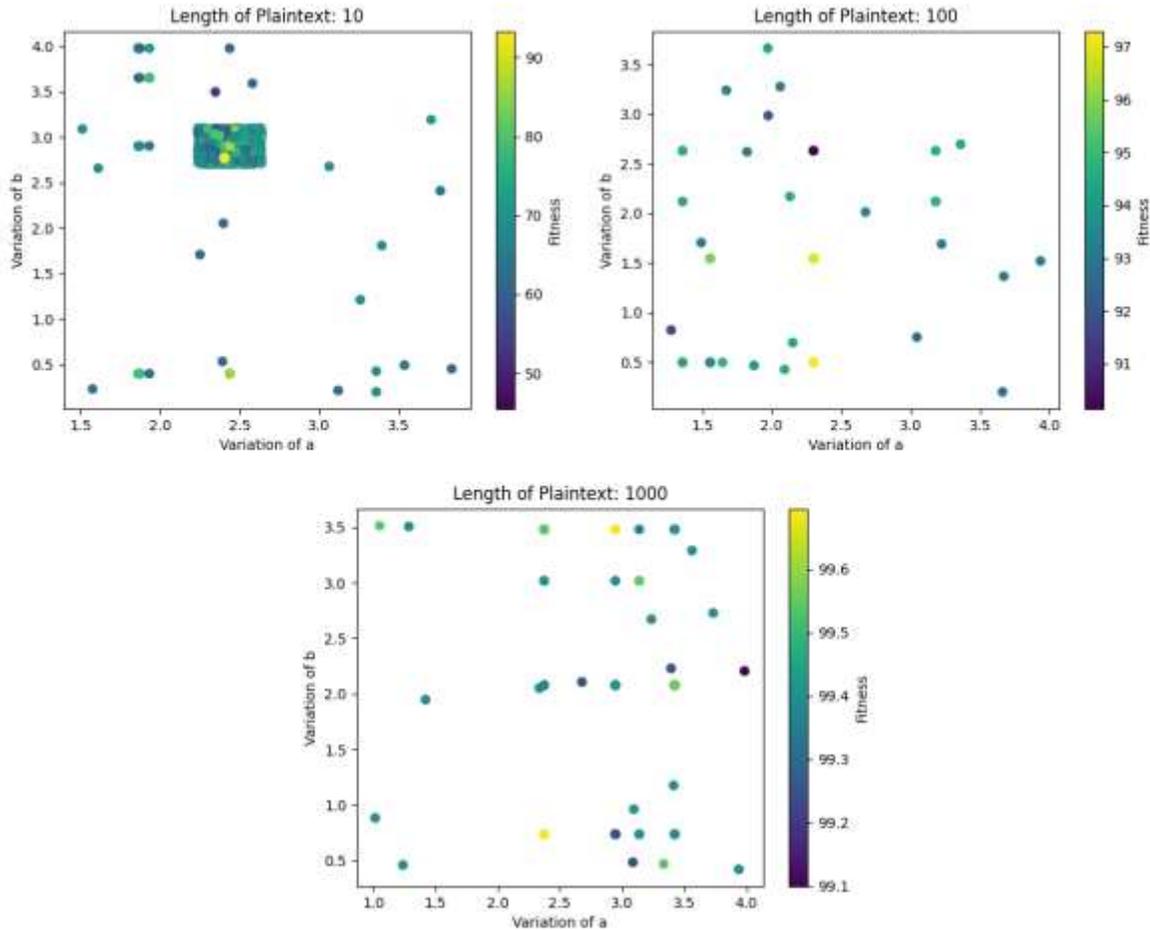

Fig. 5. Variation of Fitness Values for Individuals with different Coefficients for Fixed Plaintext Size.

Additionally, a secondary analysis was performed to study how changing the length of plaintext affects the number of generations till termination as well as the maximum fitness scores achieved. Table 1 depicts these variations. The population size for each generation was fixed to be 20 individuals. Note that upon increasing the population parameter, the range of variations covered by individuals in each generation increases significantly, thereby reducing the number of generations required and hence converging to the optimum set of parameters in a shorter runtime. All codes relevant to the analysis presented in this paper have been compiled into a GitHub repository [20].

TABLE I. Effect of Plaintext Length on Evolution Attributes

| S.No. | Length of Plaintext | No. of Generations | Max Fitness |
|-------|---------------------|--------------------|-------------|
| 1. | 10 | 234 | 92.8571 |
| 2. | 50 | 4 | 94.5946 |
| 3. | 100 | 6 | 97.2603 |
| 4. | 300 | 5 | 98.9744 |
| 5. | 700 | 2 | 99.3658 |
| 6. | 1000 | 4 | 99.7037 |

## VII. Key Space Analysis

The key space of any cryptosystem should be very large to provide an extremely vast range of possible key combinations for an attacker to access the plaintext. For the proposed system, the keys being used for decryption are the $a$ and $b$ parameters along with the $x_0$ and $y_0$ initial values with selected ranges $(1,4)$, $(0.1,4)$, $(0,1)$ and $(0,1)$ respectively. Both $a$ and $b$ show a precision of $10^{-15}$, i.e. a variation in their values by an amount as small as $10^{-15}$ will not allow decryption to occur, while the precisions shown by both $x_0$ and $y_0$ are $10^{-16}$. Thus, the size of the entire possible key space is:

$$3 \times 10^{15} \times 3.9 \times 10^{15} \times 1 \times 10^{16} \times 1 \times 10^{16}$$
$$= 1.17 \times 10^{63} \gg 2^{128}$$

As a result, the key space provided by the proposed algorithm is large enough to resist brute force attacks.

## VIII. Future Scope

This paper presents the idea of using genetic algorithms to obtain optimized values of the parameters that are used in the chaotic function. As a proof of concept, we have used a basic XOR trapdoor. The proposed system can be further upgraded by implementing a more complex trapdoor or by using this algorithm as an optimizing precursor to existing algorithms such as AES, DES, 3DES, RSA, etc. Further work can be done in selecting an improved class of genetic algorithms that can produce better results.

## IX. CONCLUSION

This paper inspects a method that makes use of concepts stemming from genetic algorithms to optimize a novel $2D$ hybrid chaotic map. The proposed map demonstrates uniform chaotic behavior for a wide range of key parameters thereby allowing the genetic algorithm to explore various key possibilities resulting in optimal encryption for every plaintext. This technique can be used to generate keys for other established encryption schemes as well that use bit operations in the trapdoor function. The key generation system proposed in this paper is also resistant to brute force attacks thereby ensuring that it takes an infeasible amount of time and key permutations for an unauthorized interceptor to gain access owing to the large key space. The method introduced in this paper enhances the security by increasing the entropy in the key using algorithms that are analogous to evolutionary trends observed in nature.